# On the Evolution of Seyfert galaxies, BL Lacertae objects and Flat-spectrum radio quasars

**Evaristus U. Iyida[*1.2], Christian I. Eze[1.2], Finbarr C. Odo[1.2]**

[1]Astronomy and Astrophysics Research Group, Department of Physics and Astronomy, University of Nigeria
[2]Department of Physics and Astronomy, Faculty of Physical sciences, University of Nigeria, Nsukka, Nigeria

*email: uzochukwu.iyida@unn.edu.ng

## Abstract

The concept of evolutionary sequence of jetted active galactic nuclei (AGNs) has been challenged in the last few decades since AGN subclasses are considered to be different due to their viewing angle. In this paper, we collected a sample of 1108 blazars (472 flat-spectrum radio quasars, FSRQs and 636 BL Lacertae objects, BL Lacs) and 120 Seyfert galaxies (SGs) with available redshifts and spectral properties in radio, optical, X-ray and γ-ray bands in order to compare the properties of SGs and the blazar subclasses and also explore their possible unification through evolution. It is found that the ratio of the relative difference of SGs and the blazar subclasses are approximately the same implying that they have evolutionary relationship. We discovered from the results of a two-dimensional Kolmogorov-Smirnov (*K*-*S*) test, that the probabilities (*p*) for the composite spectral indices: optical-X-ray ($\alpha_{OX}$), radio-X-ray ($\alpha_{RX}$), radio-optical ($\alpha_{RO}$), and X-ray-γ-ray ($\alpha_{X\gamma}$) to come from the same parent population is *p* ar less than 0.05, implying that the null hypothesis cannot be rejected, thus, supports SGs – BL Lacs – FSRQs evolutionary sequence. Simple linear regression analyses of our data yield significant anti-correlations (*r greater than* - 0.50) between $\alpha_{RO} - \alpha_{XY}$, $\alpha_{RX} - \alpha_{XY}$ and $\alpha_{RO} - \alpha_{OX}$ for the whole and individual subsamples. The correlation turns positive (*r greater than* 0.50) on $\alpha_{RO} - \alpha_{RX}$ data. These results agree with the prediction that SGs have evolutionary relationship with the blazar subclasses.

## 1 Introduction

Active Galactic Nuclei (AGNs) are the most energetic, luminous objects in the universe with supermassive black holes lurking at their centres. The study of evolution of AGNs is very important in understanding about the host galaxies and star formation rate within the universe (AGNs/galaxy co-evolution). The properties of AGNs can be described at radio frequency by a parameter called radio-loudness factor $R$, defined as the ratio of 5 GHz radio flux to that at the optical band (Kellermann et al. 1989; Stocke et al. 1992). AGNs with $R \leq 10$ are termed radio-quiet sources while those with $R > 10$ are radio-loud AGNs (Xu et al. 1999; Zhang et al. 2021). Also, AGNs are called either jetted or non-jetted depending on the strength of the ultra-relativistic jets. For most radio-loud AGNs, the jets are very strong while few radio-quiet AGNs have strong/weak jets (Tarchi et al. 2011; Doi et al. 2013). Seyfert galaxies (SGs) are special, jetted radio-quiet AGNs with bright, star-like nuclei and emission lines that cover wide range of ionization stages. They have optical spectra that show strong Fe II multiplets, which are signs that the broad line region (BLR) and the accretion disk are directly visible as observed in radio-loud AGNs. SGs are mainly classified into type 1 and type 2 (SG1 and SG2 respectively). While the spectra of SG2 have single set of relatively narrow emission lines with full width at half-maximum (FWHM) in the range of 300 to 1000 km $s^{-1}$, those of SG1 have broad emission lines in the optical spectra, FWHM $\geq$ 1000 km $s^{-1}$ (Antonucci 1993). SGs have been observed to display some unique properties such as rapid, large amplitude variability, compact radio cores, high brightness temperature, spectral variability, increased continuum emission and flat X-ray/γ-ray spectra (Yuan et al. 2008; Abdo et al. 2009; Foschini et al. 2017).

Within the past few decades, evidence has shown that the two classes of SGs are intrinsically the same type of objects though appear different mostly due to the alignment of the obfuscating materials that have toroidal geometry within the AGNs (see, Antonucci 1993; Urry and Padovani 1995). In the standard view of unification scheme for SGs, the SG2 are observed if the dusty torus intercepts an observer's line of sight (edge on view), thus, blocking the direct view of the BLR and accreting black hole while in SG2, the observer's line of sight is further away from the obscuring torus (pole on view), such that the BLR and accreting black hole are directly visible (Antonucci and Miller 1985; Urry and Padovani 1995). Similarly, some key studies support the fact that SGs1 and SGs2 are the same class of objects. For instance, Moran et al. (2000) proposed that SG1 and SG2 have the same broad emission lines in the polarized optical and infrared spectra. Also, Gallimore et al. (2010) found that SGs1 show silicate emissions on average while their SGs2 counterparts have silicate absorption, which is broadly compatible with the obscuring torus interpretation. Mas-Hesse et al. (1994) discovered that the distributions of the hard X-ray emission are similar for both SGs1 and SGs2. Also, Maiolino et al. (1997) and Curran (2000) found no difference in the mean ratio of far-infrared luminosities between the two classes of SGs, suggesting that they have the same amount of molecular gas, thus implying that they are intrinsically similar objects.

On the other hand, another group of jetted AGNs that display observational properties that are similar to SGs are the radio-loud blazars. These sources are the most prominent and persistent non-thermal emitters with a continuum powered by relativistic jets that are aligned with an observer's line of sight. They have continuum variations on all accessible timescales and spread across the entire electromagnetic spectrum. Blazars are categorized into two: the flat-spectrum radio quasars (FSRQs) and BL Lacertae objects (BL Lacs). The main difference between the two subclasses is based on their spectral properties. FSRQs have broad emission lines while BL Lacs are characterized by either featureless continua, or by spectra that show only absorption features (usually from the host galaxy) or weak narrow emission lines (Giommi et al. 2012).

The spectral energy distributions (SEDs) of SGs have been reported to be similar to blazars and are completely dominated by non-thermal radiation which can be expressed as a power law spectrum of the form $S_v \propto S^{-\alpha}$. (Abdo et al. 2009; Zhang et al. 2013; Sun et al. 2014; 2015; Foschini et al. 2015). In fact, Paliya et al. (2013) compared the SEDs of SGs, PKS 2004−447 and PKS 1502+036 with the blazar subclasses of 3C 454.3 and Mrk 421 and argued that their SEDs are similar, occupying the same parameter space at intermediate values. Lately, Chen et al. (2021) obtained significant anti-correlation between jet kinetic power and synchrotron peak frequency of blazar sample and that of narrow line SGs, which proposes that SGs could fit well into the blazar group. However, the most prominent feature of the SEDs of SGs and blazar subclasses are two peak structures, one extending from radio to X-ray and the other covering up to X-ray and γ-ray bands. The first peak is believed to be due to synchrotron emission while the second one is due to Inverse-Compton *IC* scattering from the relativistic jets. This has been found to be dominant in blazar astronomy due to the alignment of the jet axis close to the line of sight (Böttcher, 2007). For majority of blazars, the synchrotron peaks are usually located between the radio to X-ray frequencies (Giommi et al. 1995), while the Compton peaks are located in the γ-ray region (see, Finke, 2013). However, some BL Lacs are known to have Compton peaks in T eV γ-ray frequencies (Li et al. 2010). Consequently, the synchrotron peak energy ($\nu_{e,peak}$) takes wide range of values (from ∼0.40 eV to ∼13.00 eV at the extreme) and is a key method used to categorize individual BL Lacs. Sources with $\nu_{e,peak}$ ~0.40 eV are called low energy peaked BL Lacs (LBLs), those with $\nu_{e,peak}$ between ~0.40 and 4.00 eV are called intermediate energy peaked BL Lacs (IBLs) while those with $\nu_{e,peak}$ ~13.00 eV) are called high energy peaked BL Lacs (HBLs) (Padovani & Giommi 1995).

Several researches on jetted AGNs has been directed to the development of a scheme in which the observed properties of these sources could be explained as similar objects seen at different orientation angles to the line of sight. In this framework, all jetted AGNs are presumed to possess identical morphological structures and physical processes such that observations at different angles give rise to the diverse classes (Antonucci 1993). In the case of jetted radio-loud AGNs, the parent population of FSRQs are Fanaroff-Riley type II (FR II) radio galaxies (Urry et al. 1991; Urry and Padovani 1995). This means that FSRQs could be seen if FR IIs are observed along the relativistic jets. Similarly, Fanaroff-Riley type I (FR I) radio galaxies make up the parent populations of BL Lacs. However, there are some FR I sources that are associated with FSRQs, and also FR IIs that appear like BL Lacs (Ghisellini et al. 1993). Notably, there is a proposal that blazar subclasses can be arranged in a sequence from HBLs to FSRQs through LBLs and IBLs in order of decreasing synchrotron peak energies and increasing source power called blazar sequence (e.g. Fossati et al. 1998, Ghseillini et al. 1998). The most significant of this proposal is the anti-correlation between the synchrotron peak frequency and the synchrotron peak luminosity of the blazar subclasses. This phenomenon has been well studied in blazar astronomy (Fossati et al., 1998; Ghisellini et al. 1998; Ghisellini and Tavecchio, 2008; Abdo et al. 2010; Pei et al. 2020; Iyida et al. 2021a).

However, some investigations in the literature have suggested possible unification of SGs and the blazar subclasses (Chen and Gu 2019; Iyida et al. 2021c). These studies appear to offer meaningful suggestions that SGs which harbor strong relativistic jets with small black hole mass have the tendency to evolve to BL Lacs or the conventional FSRQs (Abdo et al. 2009; 2010; Foschini 2017). Specifically, a number of outstanding SGs has been observed to show blazar properties, for instance, RXJ 16290+4007 was found to show rapid variability in the radio, optical and X-ray bands, and has been observed to emit in T eV $\gamma$-ray band (see, Schwope et al. 2000; Falcone et al. 2004; Grupe et al. 2004). Whereas the sample of blazars are known to be classical radio-loud objects, the SGs are understood to possess ultra-relativistic jets with large-scale radio structures that are comparable to blazars (Liu and Zhang 2002; Mathur et al. 2012; Chen and Gu 2019). Following this line of argument, this forms a turning point for the quest to look at how radio-quiet SGs evolve into blazar subclasses based on the presence of relativistic jets. In particular, Iyida et al. (2021b) alluded that SGs can be unified with the traditional radio-loud blazar subclasses by simply invoking their broadband emission properties.

However, detailed multi-wavelength studies, especially since the launch of the *Fermi* Large Area Telescope (*Fermi*-LAT), have detected numerous jetted AGNs sources ranging from the low energy radio to the high energy γ-ray bands (Yuan et al. 2008; Abdo et al. 2009; Foschini et al., 2011). Thus, this has increased our understanding of these sources and offering new prospects of studying their multi-wave properties (see, Ajello et al. 2020; Abdolahhi et al. 2020). Correlation analyses in different wavebands are important for us to understand broadband emissions of these sources. For example, if radio emission is produced by high-energy electrons in the jet through synchrotron radiation, it is likely that these electrons could contribute part of the γ-rays due to inverse Compton scattering. Several studies have been carried out on the spectral properties of these sources. Recent analyses of blazar samples suggest a close relationship between multi-wavelength properties of blazars and their unification scheme (Savolainen et al. 2010; Iyida et al. 2022). Similarly, investigations of γ-ray data appear to provide evidence that orientation-based unification scheme is essential in explaining the variation of γ-ray emissions from blazars (e.g. Chen et al. 2016; Odo and Aroh, 2020). This is supported by many correlations between the composite spectral indices of FSRQs and BL Lacs in multiple wavebands (see, Fan et al. 2016; Yang et al. 2018; Ouyang et al. 2021; Iyida et al. 2022). In particular, the fact that the spectral flux of these sources has been found to depend strongly on the viewing angle, suggests a systematic trend in their variation from the low energy to the high energy γ-ray regions which can represent a form of unification scheme for the jetted AGNs. Thus, the new proposal should hypothetically, embrace these radio-quiet SGs as jetted counterparts of traditional radio-loud AGNs or rather, a part of a larger AGNs class that are observed under particular geometry and inclination (Singh and Chand, 2018; Iyida et al. 2021b).

Subsequently, since SGs and blazars have fully developed relativistic jets and possess similar observational properties with some scaling factors, it poses many interesting questions concerning their relationship and this challenges our current study. In this paper, we aim to statistically investigate the evolutionary relationship between the composite spectral indices of SGs and blazar subclasses. To achieve this goal, we exploited a large sample of multi-wave properties of these sources.

## 2. Selection of Samples

The data samples used in this paper are based on the *Fermi*-LAT data catalogue. From the third (3FGL) catalogue, Yang et al. (2018) compiled the composite spectral indices of 1108 optically identified blazars consisting of 472 FSRQs and 636 BL Lacs (164 LBLs, 179 IBLs and 293 HBLs). We selected redshift and the composite spectral indices of these sources in optical-X-ray ($\alpha_{OX}$), radio-X-ray ($\alpha_{RX}$), radio-optical ($\alpha_{RO}$), and X-ray-γ-ray ($\alpha_{X\gamma}$). For the SGs, we selected 120 sources with observed redshift and monochromatic luminosities ($L_m$) at $1.40 \times 10^{9}$ Hz in radio, $2.52 \times 10^{17}$ Hz in X-ray and $2.52 \times 10^{23}$ Hz in γ-ray bands from the second *Fermi*-LAT (2FGL) as published by Ackermann et al. (2012) and their flux densities calculated using $S = \frac{L_m}{4\pi d_L^2}$, where $d_L$ is the luminosity distance given as $d_L = H_o^{-1} \int_0^z \left[ (1+z)^2 (1+\Omega_m z) - z(2+z)\Omega_\Lambda \right]^{-1/2} dz$. In the optical band, the magnitude of the flux density of SGs with frequency of 4.68 × $10^{14}$ Hz was obtained from the NASA/IPAC Extragalactic Database (NED; http://ned.ipac.caltech.edu/). The composite spectral indices ($\alpha_{ij}$) of SGs between two wavebands was computed using

$$\text{composite spectral indices } (\alpha_{ij}) = -\frac{(\log S_i - \log S_j).F}{\log v_i - \log v_j} \quad (1)$$

where *i* and *j* signify two arbitrary wavebands, *S* is the monochromatic flux densities of the objects at the observing frequencies *v* while *F* is the *k* correction factor. Throughout the paper, we assume a flat Λ-CDM cosmological model with $H_0$ = 73.40 km $s^{-1}$ $Mpc^{-1}$, $\Omega_{matter}$ = 0.30 and $\Omega_{vacuum}$ = 0.70. All relevant data are adjusted based on this concordance cosmology. For the statistical analyses, we used python and MatLab programming softwares to deduce the level of relationship among the spectral properties of our samples.

## 3. Results

### 3.1 Calculation of average values

For the purpose of investigating possible connections between SGs and blazar subclasses, we calculated the average values of their composite spectral indices ($\alpha_{X\gamma}$, $\alpha_{RX}$, $\alpha_{OX}$ and $\alpha_{RO}$). The results of the average values are shown in Table 1. **We discovered from the table that the average values of SGs are approximately the same with the BL Lac subclasses (IBLs and LBLs) which suggests a fundamental similarity between SGs and these subclasses of BL Lacs. This is a strong indication that these subclasses of AGNs may be related by evolution.**

**Table 1: Average values of composite spectral indices of SGs, FSRQs and BL Lac subclasses**

| **Composite spectral indices** | ***SGs*** | ***FSRQs*** | ***HBLs*** | ***IBLs*** | **LBLs** |
|---|---|---|---|---|---|
| $\alpha_{XY}$ | 0.83 ± 0.02 | 0.62 ± 0.08 | 1.12 ± 0.06 | 0.77 ± 0.07 | 0.78 ± 0.05 |
| $\alpha_{RX}$ | 0.79 ± 0.06 | 0.97 ± 0.04 | 0.64 ± 0.08 | 0.82 ± 0.05 | 0.81 ± 0.06 |
| $\alpha_{OX}$ | 1.34 ± 0.05 | 1.25 ± 0.07 | 1.50 ± 0.05 | 1.61 ± 0.03 | 1.63 ± 0.07 |
| $\alpha_{RO}$ | 0.52 ± 0.03 | 0.79 ± 0.01 | 0.38 ± 0.05 | 0.51 ± 0.03 | 0.53 ± 0.04 |

### 3.2 Comparison of the average values of the composite spectral indices

The calculated average values of $\alpha_{X\gamma}$ $\alpha_{RX}$, $\alpha_{OX}$, and $\alpha_{RO}$ of SGs were compared with those of FSRQs, HBLs, IBLs and LBLs using the relative difference defined (see, Yang et al. 2018) as

$$R_{ab} = \frac{|\alpha_{ij,a} - \alpha_{ij,b}|}{\alpha_{ij,B}} \times 100\% \qquad (2)$$

where $\alpha_{ij,a}$ and $\alpha_{ij,b}$ are the average values of the composite spectral indices of SGs and the blazar subclasses respectively. Using the results of the average values in table 1, the relative difference was obtained and the results given in Table 2. Furthermore, the ratio of the relative difference of SGs and the blazar subclasses was calculated and also given in the table 2. **We discovered from the ratio of the relative difference that SGs and blazar subclasses (IBLs and LBLs) have approximately the same values, indicating that they have common emission conditions, thus, indicating evolutionary relationship.**

**Table 2: Comparison of average values of the composite spectral indices of our samples**

| **Composite spectral indices** | $R_{SF}$ | $R_{SH}$ | $R_{SI}$ | $R_{SL}$ | $\frac{R_{SI}}{R_{SF}}$ | $\frac{R_{SL}}{R_{SF}}$ |
|---|---|---|---|---|---|---|
| $\alpha_{X\gamma}$ | 33.87 | 23.89 | 14.00 | 12.16 | 0.41 | 0.36 |
| $\alpha_{RX}$ | 23.71 | 15.62 | 15.90 | 16.85 | 0.67 | 0.71 |
| $\alpha_{OX}$ | 11.02 | 19.33 | 24.84 | 25.76 | 2.25 | 2.34 |
| $\alpha_{RO}$ | 34.17 | 36.84 | 9.60 | 11.50 | 0.28 | 0.33 |

### 3.3 Distributions of composite spectral indices

We used a two-dimensional Kolmogorov-Smirnov (*K-S*) test with a 5% significance level (probability value $p < 0.05$) to investigate the evolutionary sequence of SGs and blazar subclasses. The cumulative distribution functions of $\alpha_{X\gamma}$ $\alpha_{RX}$, $\alpha_{OX}$, and $\alpha_{RO}$ are shown in Figures 1 (a-d) respectively while the *K-S* test results are given in Table 3. In the table, column (1) gives the composite spectral index used for the *K-S* test, column (2) gives SGs and the blazar subclass used for the *K-S* test, column (3) gives the sample size ($\boldsymbol{N_a}$) of SGs, column (4) gives the sample size ($\boldsymbol{N_b}$) of blazar subclass, column (5) gives the separation distance between the cumulative probabilities of SGs and blazar subclass in the *K-S* test, column (6) gives the composite spectral index corresponding to the maximum difference of the cumulative probabilities and column (7) gives the probabilities for SGs and blazar subclasses to come from the same parent population in the distributions. We discovered from the table that the probabilities for the composite spectral indices of SGs and blazar subclasses to come from the same parent distribution are far less than 0.05, implying SGs – BL Lacs – FSRQs evolutionary sequence. This plausibly means that SGs are born as young AGN subsets (young), metamorphose through different forms of BL Lacs (adult) into FSRQs (old).

**Table: 3 The *K-S* test results of the composite spectral indices of SGs, FSRQs and BL Lac subclasses**

| **Composite spectral parameter** | ***Subsamples*** | $N_a$ | $N_b$ | $d_{max}$ | $\alpha_{d\max}$ | $p$ |
|---|---|---|---|---|---|---|
| $\alpha_{XY}$ | SGs – HBLs | 120 | 293 | 0.85 | 0.42 | $3.09\times10^{-8}$ |
| | SGs – LBLs | 120 | 164 | 0.38 | 0.17 | $2.32\times10^{-6}$ |
| | SGs - IBLs | 120 | 179 | 0.31 | 0.15 | $7.96\times10^{-9}$ |
| | SGs - FSRQs | 120 | 472 | 0.46 | 0.42 | $9.02\times10^{-5}$ |
| $\alpha_{rx}$ | SGs – HBLs | 120 | 293 | 0.63 | 0.12 | $1.02\times10^{-5}$ |
| | SGs – LBLs | 120 | 164 | 0.75 | 0.17 | $4.20\times10^{-7}$ |
| | SGs - IBLs | 120 | 179 | 0.68 | 0.15 | $2.87\times10^{-5}$ |
| | SGs - FSRQs | 120 | 472 | 0.97 | 0.22 | $4.84\times10^{-6}$ |
| $\alpha_{OX}$ | SGs – HBLs | 120 | 293 | 0.79 | 0.21 | $2.01\times10^{-8}$ |
| | SGs – LBLs | 120 | 164 | 0.94 | 0.39 | $6.42\times10^{-5}$ |
| | SGs - IBLs | 120 | 179 | 0.91 | 0.37 | $4.66\times10^{-6}$ |
| | SGs - FSRQs | 120 | 472 | 0.14 | 0.10 | $7.04\times10^{-8}$ |
| $\alpha_{RO}$ | SGs – HBLs | 120 | 293 | 0.81 | 0.23 | $3.93\times10^{-9}$ |
| | SGs – LBLs | 120 | 164 | 0.20 | 0.19 | $1.82\times10^{-7}$ |
| | SGs - IBLs | 120 | 179 | 0.11 | 0.10 | $2.89\times10^{-9}$ |
| | SGs - FSRQs | 120 | 472 | 0.94 | 0.45 | $3.07\times10^{-5}$ |

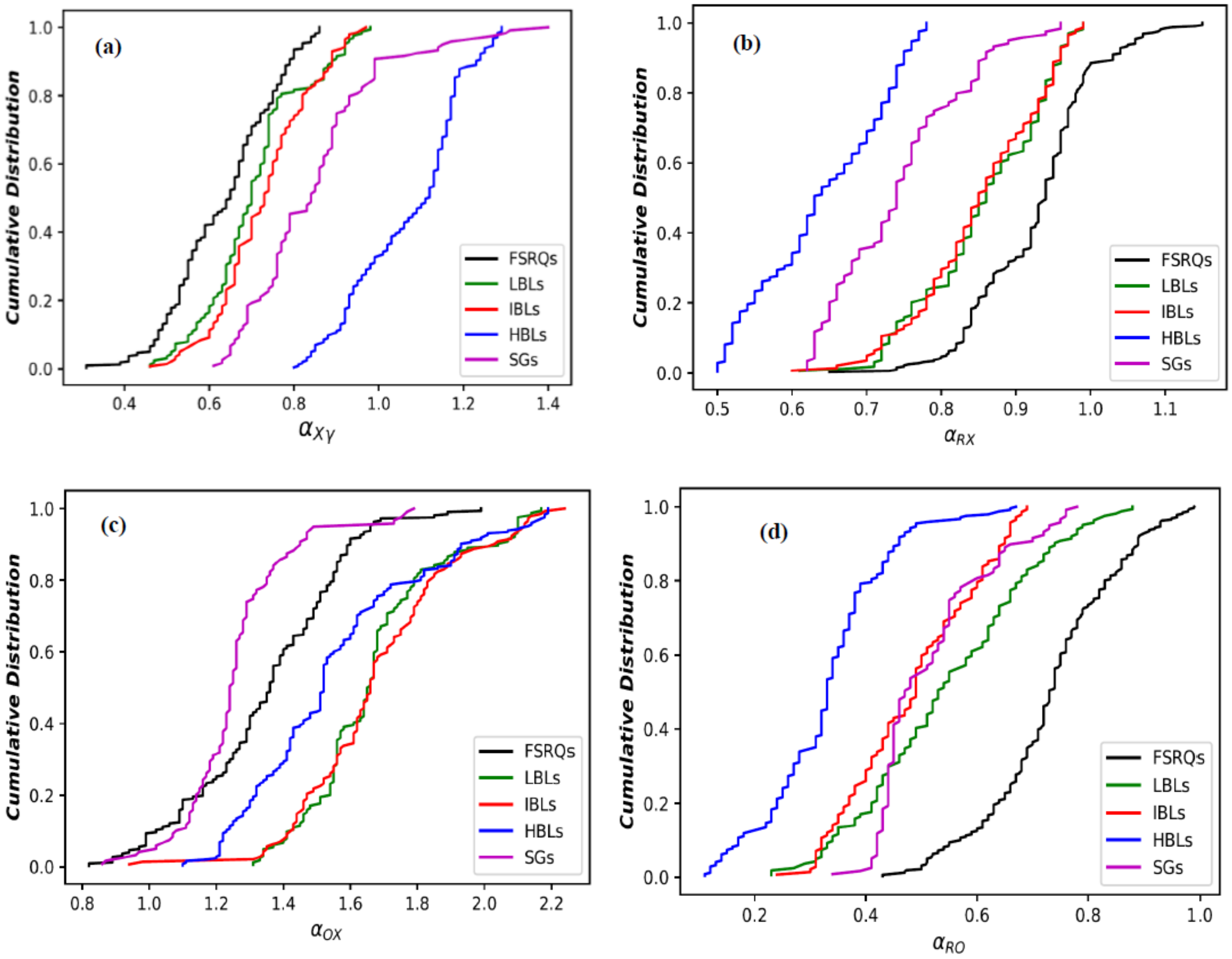

**Figure 1: Cumulative distribution function of (a) $\alpha_{X\gamma}$ (b) $\alpha_{RX}$, (c) $\alpha_{OX}$ (d) $\alpha_{RO}$ of our samples**

### 3.4 Comparison of composite spectral indices of SGs and blazar subclasses

We compare using histogram distributions, the composite spectral indices of SGs, FSRQs and BL Lac subclasses in order to ascertain their level of relationship and the possibility of SGs – BL Lacs - FSRQs evolution. The distributions of $\alpha_{XY}$ and $\alpha_{RX}$ of FSRQs, SGs and BL Lac subclasses are shown in Figures 2a and 2b respectively. It is clear from Figure 2a that FSRQs have the least values of $\alpha_{XY}$ while SGs tends to have larger values of $\alpha_{XY}$ in the data though they cover similar ranges and overlap at intermediate values. Likewise, SGs and BL Lacs have unimodal configuration while FSRQs have multimodal configuration. In the case of $\alpha_{RX}$, FSRQs range from 0.65 to 1.15 with a single peak at 0.97. However, SGs and BL Lacs range from 0.64 to 2.18 and 0.50 to 0.99 respectively with LBLs, IBLs and HBLs having peaks at different values. Further statistical analysis done on our samples using *Jarque-Bera* test (see, Jarque & Bera, 1980) reveals that FSRQs, SGs and BL Lac subclasses are not nicely fitted to normal distribution.

Meanwhile, it is observed that the distributions of $\alpha_{XY}$ and $\alpha_{RX}$ are continuous with no distinct dichotomy among FSRQs, SGs and BL Lac subclasses in such a way that signifies the continuously evolving scheme of SGs – BL Lacs – FSRQs, thus, the populations of the samples are inherently related.

We show in Figures 2c and 2d, the distributions of our samples in both $\alpha_{OX}$ and $\alpha_{RO}$. Obviously BL Lacs on average, have the tendency to possess higher values of $\alpha_{OX}$ than the FSRQs and SGs. However, FSRQs, SGs and BL Lacs are distributed in unimodal configurations. This is however, not the same for $\alpha_{RO}$ where FSRQs tends to have larger values while BL lac are displaced to the lowest values of $\alpha_{RO}$. **However, the distributions of $\alpha_{OX}$ and $\alpha_{RO}$ are continuously arranged in the sequence, SGs – BL Lacs – FSRQs with no obvious division suggesting that these sources are intrinsically similar and the scaling could arise due to different stages of evolution of the sources.**

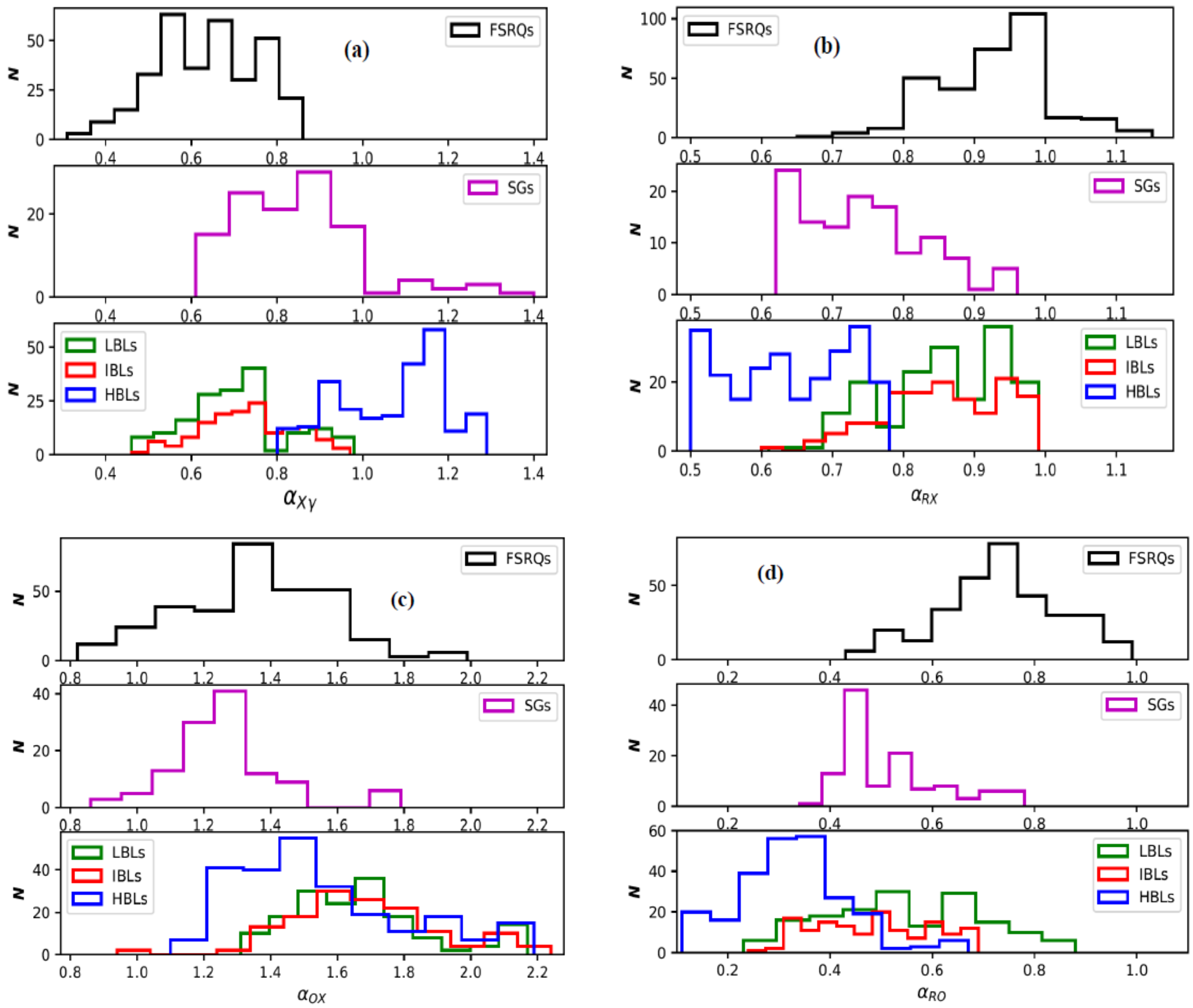

**Figure 2: Histogram showing comparison of (a) $\alpha_{X\gamma}$ (b) $\alpha_{RX}$, (c) $\alpha_{OX}$, (d) $\alpha_{RO}$ of our samples**

In order to further investigate the evolutionary relationship among SGs, BL Lacs and FSRQs, we show in Figure 3a, the distribution of the spectroscopic redshift of our samples. From the Figure, the range of redshifts of FSRQs and BL Lacs is $0.068 < \log(1+z) < 0.52$ and $0.03 < \log(1+z) < 0.29$ respectively while the range of SGs is $0.001 < \log(1+z) < 0.03$ which is about 10 % less than FSRQs and BL Lacs.  A *K-S* test was carried out on the redshift data. Our result generally shows that, there is approximately zero probability ($p = 2 \times 10^{-24}$) that there is a fundamental difference between the underlying distributions of SGs, BL Lacs and FSRQs in redshift. The cumulative probability function is shown in Figure 3(b). From the distribution and the *K-S* test results, we find that the mean values of the redshift of our samples follow the sequence

$$\langle \log(1+z) \rangle_{SGs} < \langle \log(1+z) \rangle|_{HBLs} < \langle \log(1+z) \rangle|_{IBLs} < \langle \log(1+z) \rangle|_{LBLs} < \langle \log(1+z) \rangle|_{FSRQs}$$

indicative of SGs – BL Lacs – FSRQs evolution. This result proposes the different stages of cosmological evolution of the jetted AGNs: SGs with the lowest redshift grow into BL Lacs, and then FSRQs at high redshifts (young – adult – old), thus, signifying some evolutionary sequence of SGs and the blazars types.

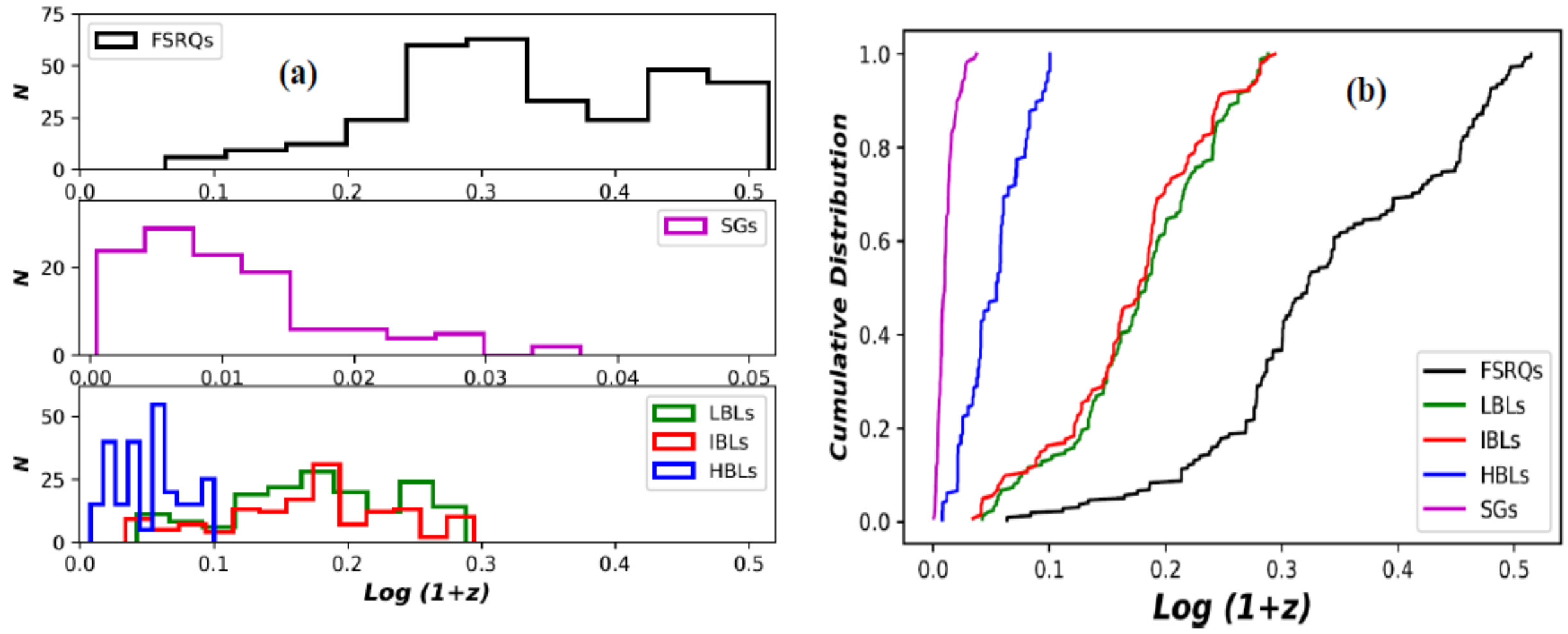

**Figure 3: Histogram showing comparison of logarithmic of (a) (1+z) of SGs, BL Lacs and FSRQs (b) cumulative distribution function of (1+z) of the samples**

### 3.5 Composite spectral indices and SGs – BL Lacs – FSRQs evolution

The correlations among source parameters are not only necessary for studying their unification scheme via orientation but also useful in understanding their pattern of evolution. In this subsection, we analyzed whether the correlations found among the parameters of SGs, and blazar subclasses are in agreement with the evolutionary sequence using the Pearson correlation theory.

The Pearson's regression equation is expressed as $y = (a \pm \Delta a)x + (b \pm \Delta b)$, with $a$ being the slope while $b$ is the intercept. We computed the Pearson product-moment correlation coefficient ($r$) for the whole and individual subsamples using the expression given as (Press 1994; Pavlidou et al. 2012):

$$r = \frac{\sum (x_i - \bar{x})(y_i - \bar{y})}{\sqrt{\sum (x_i - \bar{x})(\sqrt{\sum y_i - \bar{y}})}} \qquad (3)$$

$\bar{x}$ and $\bar{y}$ are the average values of $x_i$ and $y_i$. We show respectively in Figures 4a and 4b, the scatter plots of $\alpha_{RO}$ and $\alpha_{RX}$ as a function of $\alpha_{XY}$. It is clear from the figures that SGs, LBLs and IBLs overlap in a sense that is consistent with the unified scheme via blazar sequence. There is a clear anti-correlation of $\alpha_{RO} - \alpha_{XY}$ and $\alpha_{RX} - \alpha_{XY}$ data in each plot. However, it can be seen in the figures that while SGs evolve into IBLs and LBLs, there is a fundamental difference between SGs and HBLs. Therefore, it could be argued that the observed dichotomy between SGs, and HBLs in $\alpha_{RO} - \alpha_{XY}$ and $\alpha_{RX} - \alpha_{XY}$ planes is as a result of the differences in their intrinsic properties that vary in sequence across the samples. The results of the linear regression analyses of the samples are shown in table 4.

Conversely, while the $\alpha_{RO}$ is strongly anti-correlated with $\alpha_{OX}$ for SGs, FSRQs and BL Lacs subclasses as shown in the scatter plot in Fig. 4(c), that of $\alpha_{RO} - \alpha_{RX}$ is positively correlated as shown in Fig. 4(d). It can be observed that the spread in $\alpha_{RO}$ for different values of $\alpha_{OX}$ and $\alpha_{RX}$ is significantly the same for SGs, FSRQs and BL subclasses. Also, the connections between SGs and blazar subclasses (IBLs, LBLs) are still prominent on the $\alpha_{RO} - \alpha_{OX}$ and $\alpha_{RO} - \alpha_{RX}$ planes. This again suggests that SGs have evolutionary relationship with IBLs, LBLs and FSRQs through composite spectral indices. The results of regression analysis of the subsamples are shown in table 4. The slope $a$, intercept ***b***, correlation coefficient $r$ and chance probability $p$ and their errors are all listed in the table. The strong correlations imply that similar effects are responsible for variations in the composite spectral indices of SGs, BL Lacs and FSRQs and SGs are the youngest subset of jetted AGNs.

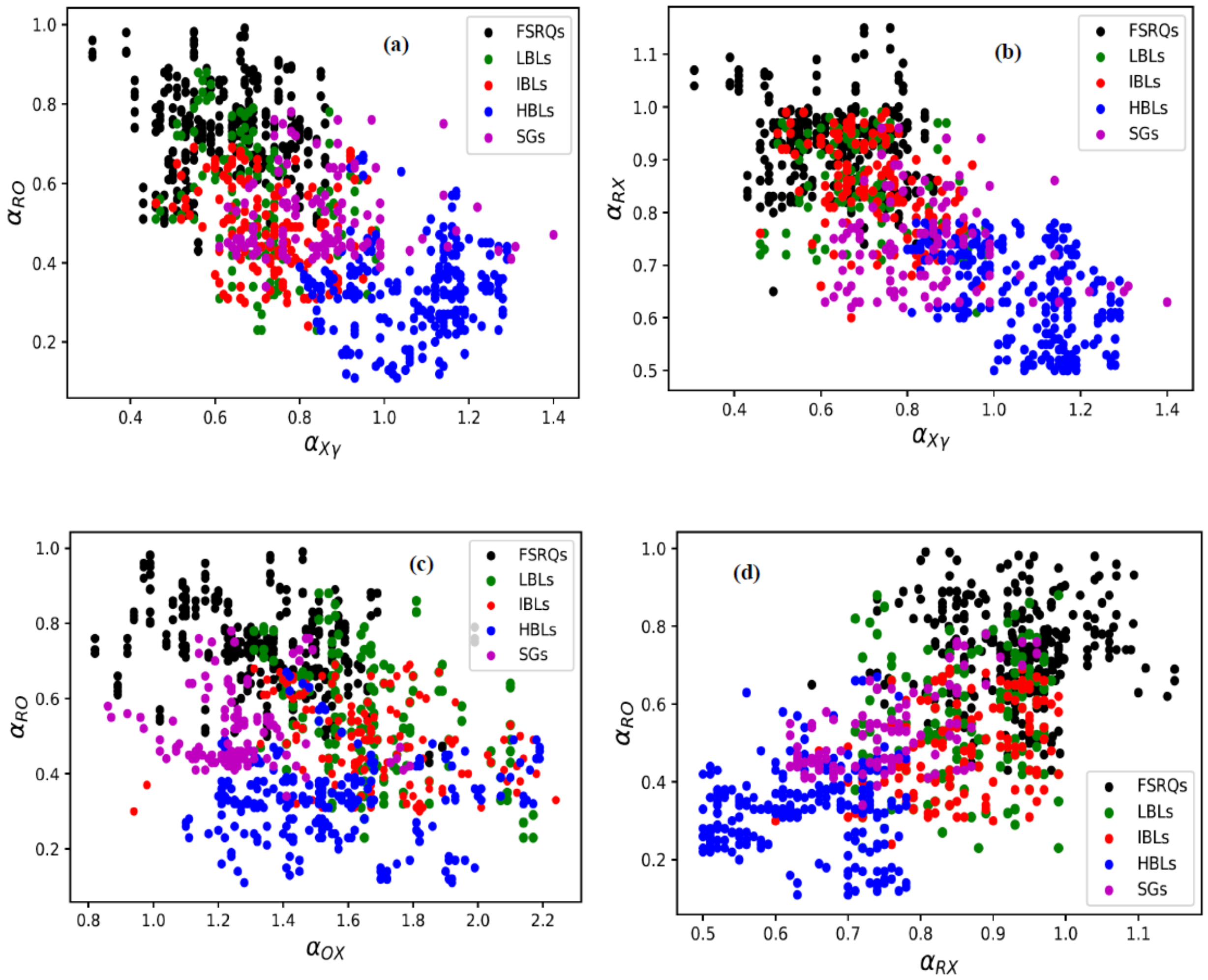

**Figure 4: scatter plots of (a) $\alpha_{RO}$ - $\alpha_{XY}$ (b) $\alpha_{RX}$ - $\alpha_{XY}$ (c) $\alpha_{RO}$ - $\alpha_{OX}$ (d) $\alpha_{RO}$ - $\alpha_{RX}$ of our samples**

**Table 4: Results of linear regression analyses of the composite spectral indices of our samples**

| ***plots*** | **Subsamples** | ***a*** | ***Δa*** | ***b*** | ***Δb*** | *r* | ***p*** |
|---|---|---|---|---|---|---|---|
| $\alpha_{RO}$ - $\alpha_{XY}$ | Whole | 1.30 | 0.04 | 0.89 | 0.02 | - 0.79 | $2.90\times10^{-4}$ |
| | SGs | 0.59 | 0.03 | 0.74 | 0.03 | -0.51 | $2.83\times10^{-6}$ |
| | FSRQs | 0.63 | 0.07 | 0.73 | 0.05 | -0.53 | $2.89\times10^{-5}$ |
| | HBLs | 0.42 | 0.04 | 0.62 | 0.03 | -0.50 | $5.07\times10^{-6}$ |
| | IBLs | 0.42 | 0.05 | 0.64 | 0.05 | -0.49 | $1.93\times10^{-9}$ |
| | LBLs | 0.57 | 0.02 | 0.61 | 0.04 | -0.52 | $5.82\times10^{-7}$ |
| $\alpha_{RX}$ - $\alpha_{XY}$ | Whole | 0.94 | 0.04 | 1.06 | 0.02 | -0.84 | $4.89\times10^{-9}$ |
| | SGs | 0.24 | 0.05 | 0.91 | 0.02 | -0.54 | $3.07\times10^{-5}$ |
| | FSRQs | 0.42 | 0.07 | 0.87 | 0.05 | -0.46 | $4.93\times10^{-5}$ |
| | HBLs | 0.30 | 0.02 | 0.92 | 0.04 | -0.64 | $1.82\times10^{-7}$ |
| | IBLs | 0.27 | 0.05 | 0.88 | 0.05 | -0.53 | $2.29\times10^{-6}$ |
| | LBLs | 0.41 | 0.06 | 0.86 | 0.03 | -0.50 | $1.37\times10^{-5}$ |
| $\alpha_{RO}$ - $\alpha_{OX}$ | Whole | 0.74 | 0.03 | 0.81 | 0.04 | -0.69 | $1.99\times10^{-6}$ |
| | SGs | 0.34 | 0.05 | 0.72 | 0.05 | -0.52 | $1.85\times10^{-8}$ |
| | FSRQs | 0.21 | 0.04 | 0.98 | 0.03 | -0.54 | $2.09\times10^{-6}$ |
| | HBLs | 0.18 | 0.02 | 0.54 | 0.04 | -0.45 | $3.80\times10^{-5}$ |
| | IBLs | 0.33 | 0.03 | 0.67 | 0.03 | -0.49 | $1.13\times10^{-5}$ |
| | LBLs | 0.24 | 0.04 | 0.72 | 0.03 | -0.52 | $5.82\times10^{-7}$ |
| $\alpha_{RO}$ - $\alpha_{XY}$ | Whole | 0.89 | 0.03 | 0.32 | 0.02 | 0.73 | $3.89\times10^{-9}$ |
| | SGs | 0.56 | 0.03 | 0.36 | 0.05 | 0.56 | $3.07\times10^{-5}$ |
| | FSRQs | 0.34 | 0.05 | 0.53 | 0.06 | 0.47 | $2.93\times10^{-9}$ |
| | HBLs | 0.43 | 0.06 | 0.52 | 0.03 | 0.45 | $3.82\times10^{-7}$ |
| | IBLs | 0.32 | 0.06 | 0.44 | 0.05 | 0.46 | $1.89\times10^{-9}$ |
| | LBLs | 0.35 | 0.04 | 0.45 | 0.04 | 0.47 | $3.07\times10^{-5}$ |

**4. Discussion**

Blazars are distinct subclass of radio-loud AGNs whose relativistic jets are aligned close to an observer's line of sight. These sources are characterized by compact radio cores, flat radio spectra, high brightness temperature and rapid variability in multi-wave bands (Urry and Padovani 1995). These unique properties have been reportedly observed in jetted radio-quiet SGs (e.g. Yuan et al. 2008; Abdo et al. 2009; Foschini et al. 2015; 2017). Also, since SGs and blazar subclasses lie at the opposite extremes in AGNs parameter space, the discovery of their hybrids in the past few decades have drawn considerable attention about their relationship (see, Zhou et al. 2003; 2007). We have used the multi-wavelength properties of these jetted AGNs to investigate the evolutionary relationship between the SGs and their blazar counterparts in a sample of *Fermi*-LAT data catalogue. These jetted sources form an exceptional population of AGNs with similar emission processes effective under a variety of physical conditions. The blazar sequence proposes that both the FSRQs and BL Lac objects are forms of objects with the same physical process but differ in bolometric luminosity (Fossati et al. 1998; Ghisellini et al. 1998; Abdo et al. 2010). This infers that FSRQs and BL Lac subclasses are progressively misaligned subclasses of AGNs. **The new unified scheme of AGNs posits that these blazar-like galaxies called SGs are young jetted counterparts of the traditional radio-loud AGNs or a portion of a larger subclass of AGNs that are detected differently due to either different orientation with respect to the line of sight (Singh and Chand 2018) or observations at different stages of evolution (Chen and Gu 2019; Cheng et al. 2021). Therefore, these objects can supposedly be unified with the conventional blazar subclasses through evolution, orientation or blazar sequence. In this sense, there should be continuity in distributions of the properties of SGs, FSRQs and BL Lacs.**

The distributions of composite spectral indices and redshift of our samples are actually in agreement with the evolutionary scheme, as there is no dichotomy between SGs and blazar subclasses which signifies that these sources are intrinsically the same class of objects that differ only by evolution. The interpretation is that the main mechanism that is generating the multi-wave emissions in our samples are the same, though, they differ systematically among the AGNs subclasses. This indicates that there is evolutionary link among these jetted AGNs subclasses which is in agreement with the earlier proposal by Fossati et al. (1998) and Ghisellini et al. (1998) and between blazars and γ-ray narrow line Seyfert 1 galaxies (Chen et al. 2021). Notably, our results in particular show from the comparison of the composite spectral indices that jetted SGs completely overlap in values of $\alpha_{X\gamma}$ $\alpha_{RX}$, $\alpha_{OX}$, and $\alpha_{RO}$ with the blazar subclasses of FSRQs and BL Lacs. Similarly, these observations seem to suggest that SGs and blazar subclasses are the same class of objects that differ only by the orientation and location of the observer. In fact, Dahari and De Robertis (1988) and Falcke et al. (2000) have already pointed out the similarities between the nuclei of radio-quiet SGs and radio-loud AGNs suggesting that they are similar. Also, Ho et al. (1997), Nagar et al. (2000), Foschini et al. (2015) and Iyida et al. (2021c) demonstrated a continuity in overall distributions of observed properties of the SGs and traditional radio-loud AGNs. Our results support the proposal that blazar subclasses are similar to that of narrow line Seyfert 1 galaxy in jet power and accretion disk luminosity in that the formation mechanism of the jets in narrow line Seyfert 1 galaxy is similar to that of blazar subclasses (Chen and Gu 2019; Cheng et al. 2021). Therefore, if this is actually the case, it follows that the continuous distributions observed in the values of SGs and blazar subclasses are evidence that they have evolutionary relationship.

It is proposed that FR IIs are the parent populations of BL Lacs and believed to harbor powerful relativistic jets with extended radio structures that are comparable to SGs (see, Liu and Zhang, 2002; Doi et al. 2012; Mathur et al. 2012). Therefore, one can expect that, for the composite spectral indices, the distribution of SGs and BL Lacs should be from the same parent distribution. We investigated the distributions using average values and relative difference. For the SGs, the average values and relative difference in $\alpha_{X\gamma}$ $\alpha_{RX}$, $\alpha_{OX}$, and $\alpha_{RO}$ are approximately the same as for BL Lacs (IBLs, LBLs). Also, their cumulative distribution functions for $\alpha_{X\gamma}$ $\alpha_{RX}$, and $\alpha_{RO}$ are very close to that of BL Lacs. When the *K–S* tests are adopted to our samples, it is found that the probabilities for $\alpha_{X\gamma}$ $\alpha_{RX}$, $\alpha_{OX}$, and $\alpha_{RO}$ to be from the same parent distribution are far less than 0.05, implying that the null hypothesis cannot be rejected, thereby suggesting that the composite spectral indices of SGs and BL Lacs are from the same parent distribution, which supports SGs – BL Lacs – FSRQs evolution. The fact that SGs, IBLs, LBLs and FSRQs occupy nearly like sections in the $\alpha ro$ - $\alpha rx$, $\alpha_{RX}$ - $\alpha_{XY}$, $\alpha_{RO}$ - $\alpha_{OX}$ and $\alpha_{RO}$ - $\alpha_{RX}$ implies that these jetted AGN subclasses have similar evolution and histories (e.g. Urry and Padovani 1995; Pei et al. 2019). Our results show from Fig. 4d that SGs and BL Lac subclasses of IBLs and LBLs are completely mixed up while HBLs occupy distinct region in the color-color diagram, which suggests SGs,

FSRQs, IBLs and LBLs have similar composite spectral properties while HBLs have different spectral properties.

Also, the results of the current investigation show significant anti-correlation ($r \geq$ - 0.50) between the $\alpha_{RO} - \alpha_{XY}$, $\alpha_{RX} - \alpha_{XY}$ and $\alpha_{RO} - \alpha_{OX}$ for SGs, FSRQs and BL Lac subclasses. This provides more hints and evidence for the evolutionary model of SGs and blazar subclasses. On the other hand, the distributions of composite spectral indices revealed from Fig. 4 (a-c) suggests that the spectral trend of HBLs is different from that of SGs, IBLs, LBLs and FSRQs and there is a continuous trends going from FSRQs at the upper left corner to HBLs at the lower right corner through SGs, IBLs and LBLs, which is consistent with the blazar sequence. Thus, it can be deduced that similar mechanisms are responsible for their variation with some forms of scaling factors on their composite spectral indices. Thus, this sequence of variation from FSRQs, through SGs, IBLs, LBLs to HBLs is in a sense that indicates evolution. Perceptibly, SGs, FSRQs, IBLs and LBLs are almost indistinguishable and occupies the same parameter space in $\alpha_{RO} - \alpha_{XY}$, $\alpha_{RX} - \alpha_{XY}$ and $\alpha_{RO} - \alpha_{OX}$ plots as shown in Fig. 4., this space occupied by the SGs, relative to blazar subclasses, is in good agreement with evolutionary sequence, suggesting that similar mechanisms are operational in these objects. The order in which SGs, IBLs and LBLs appear in the current study is a signature that there might be a kind of relationship for these objects. Nevertheless, for the BL Lacs, the clear separation of HBLs from SGs, IBLs and LBLs on the $\alpha_{RO} - \alpha_{XY}$, $\alpha_{RX} - \alpha_{XY}$, $\alpha_{RO} - \alpha_{RX}$ and $\alpha_{RO} - \alpha_{OX}$ planes can be interpreted in terms of a fundamental difference between HBLs and other subclasses of BL Lacs.

## 5. Conclusions

We have studied the composite spectral indices of jetted SGs, and their blazar counterparts of FSRQs and BL Lac subclasses. We discovered from the ratio of relative difference that SGs and blazar subclasses have the same values which is an evidence that these sources are the same class of objects but observed differently. We used the *K-S* test to analyze the cumulative distribution functions of the composite spectral indices of SGs and blazar subclasses and found that the probabilities $p$ for the jetted SGs and blazar subclasses to come from the same parent distribution $p << 0.05$, indicating that they are fundamentally the same class of objects and thus, can be unified via evolution. There are significant anti-correlations ($r \geq$ - 0.50) between $\alpha_{RO} - \alpha_{XY}$, $\alpha_{RX} - \alpha_{XY}$ and $\alpha_{RO} - \alpha_{RX}$ data for SGs and individual subsamples of blazar. This upturns into positive correlation ($r \geq 0.50$) on the $\alpha_{RO} - \alpha_{OX}$ plane. These significant correlations imply that SGs have evolutionary relationship with the blazar subclasses.

**Data Availability.**

The data sample used in this paper is available to readers on request at uzochukwu.iyida@unn.edu.ng

**Acknowledgement**

We thank anonymous referee immensely for valuable comments and suggestions which helped us to improve the manuscript. This paper used the NASA/IPAC Extragalactic Database (NED) which is operated by the Jet Propulsion Laboratory, Caltech, under contract with the National Aeronautics and Space Administration.